\documentclass{aa}
\usepackage{txfonts}
\usepackage{graphicx}
\usepackage[latin1]{inputenc}
\begin{document}
\newcommand{\HII}{H\,{\sc{ii}}}
\newcommand{\SII}{S\,{\sc{ii}}}
\newcommand{\HI}{H\,{\sc{i}}}
\newcommand{\msol}{\hbox{\kern0.20em$M_\odot$}}

\newcommand{\frs}{\hbox{$.\!^{\rm s}$}}

\newcommand{\kms}{\hbox{\kern0.20emkm\kern0.20ems$^{-1}$}}
\newcommand{\cmmt}{\hbox{\kern0.20emcm$^{-3}$}}
\newcommand{\cmmd}{\hbox{\kern0.20emcm$^{-2}$}}
\newcommand{\pc}{\hbox{\kern0.20empc$^{2}$}}
\newcommand{\twco}{\hbox{${}^{12}$CO}}
\newcommand{\twcotwo}{\hbox{${}^{12}$CO(2-1)}}
\newcommand{\thco}{\hbox{${}^{13}$CO}}
\newcommand{\thcotwo}{\hbox{${}^{13}$CO(2-1)}}
\newcommand{\thcoone}{\hbox{${}^{13}$CO(1-0)}}
\newcommand{\ceio}{\hbox{C${}^{18}$O}}
\newcommand{\ceiotwo}{\hbox{C${}^{18}$O(2-1)}}
\newcommand{\ceioone}{\hbox{C${}^{18}$O(1-0)}}
\newcommand{\cs}{\hbox{CS}}
\newcommand{\street}{\hbox{CS(3-2)}}
\newcommand{\castor}{\hbox{CS(2-1)}}
\newcommand{\costive}{\hbox{CS(5-4)}}
\newcommand{\Acts}{\hbox{C${}^{34}$S}}
\newcommand{\cutthroat}{\hbox{C${}^{34}$S(3-2)}}
\newcommand{\cotton}{\hbox{C${}^{34}$S(2-1)}}
\newcommand{\two}{\hbox{H${}_2$}}
\newcommand{\Halpha}{\hbox{H$\alpha$}}
\newcommand{\cop}{\hbox{HCO$^{+}$}}
\newcommand{\concertgoer}{\hbox{$J=1\rightarrow 0$}}
\newcommand{\tontine}{\hbox{$J=2\rightarrow 1$}}
\newcommand{\thereto}{\hbox{$J=3\rightarrow 2$}}
\newcommand{\fortieth}{\hbox{$J=4\rightarrow 3$}}
\newcommand{\five}{\hbox{$J=5\rightarrow 4$}}

\title{Triggered massive-star formation on the borders\\
       of Galactic H\,{\Large II} regions}
\subtitle{I. A search for `collect and collapse' candidates}

\author{L.~Deharveng
        \and
        A.~Zavagno
       \and
         J.~Caplan
        }
  \authorrunning{L.~Deharveng et al.}
  \offprints{L.~Deharveng}

\institute{
      Laboratoire d'Astrophysique de Marseille, 
      2 Place Le Verrier, 13248 Marseille Cedex 4, France\\
             \email{lise.deharveng@oamp.fr}
      }

\date{Received 03/09/2004; accepted 30/11/2004 }

\abstract{
Young massive stars or clusters are often observed at the peripheries
of \HII\ regions. What triggers star formation at such locations?
Among the scenarios that have been proposed, the `collect and
collapse' process is particularly attractive because it
permits the formation of massive objects via the fragmentation of the
dense shocked layer of neutral gas surrounding the expanding ionized
zone. However, until our recent article on Sh~104, it had not been
convincingly demonstrated that this process
actually takes place. In the present paper we present our selection of
seventeen candidate regions for this process; all show high-luminosity
near-IR clusters and/or mid-IR point sources at their peripheries. The
reality of a `collect and collapse' origin of these presumably 
second-generation stars and clusters will be discussed in forthcoming 
papers, using new near-IR and millimetre observations.

   }

\titlerunning{Triggered massive-star formation: collect and collapse
candidates}
\authorrunning{L.~Deharveng et al.}

\maketitle

\noindent

\section{Introduction\label{intro}}

Statistical studies have shown that the presence of an \HII\
region contiguous to a molecular cloud has two effects on star
formation in the cloud: an increased rate of formation in general, and
increased formation of massive objects in particular. For example,
Dobashi et al.~(\cite{dob01}) investigated the luminosity of protostars
in molecular clouds, as a function of the cloud mass, in a sample of
499 clouds taken from the literature; 243 of these clouds turned out to
be associated with protostar candidates selected from the IRAS Point
Source Catalog. They showed that the protostars in clouds adjacent to
\HII\ regions are more luminous than those in the clouds away from \HII
\ regions. Finding that there are well-defined upper and lower
protostar luminosity limits, they proposed a very simple model in which
the protostar luminosity is controlled by the mass of the parental
cloud and by an external pressure exerted upon the cloud surface. The
lower and upper limits of the luminosity distribution correspond to
external pressures of zero and $\sim10^{5.5}\, k$~K~cm~$^{-3}$, where
$k$ is the Boltzmann constant. This latter figure is a
reasonable value for the pressure of the ionized gas in a classical
\HII\ region.

Detailed studies of large star-forming regions also show that
the presence of an \HII\ region favours star formation, especially of
massive objects. Karr and Martin (\cite{kar03}), in their multi-wavelength
study of the W5 \HII\ region, found that the number of star-formation events
per unit CO covering area is 4.8 times higher within the influence zone of
the \HII\ region than outside. Also, in their study of the Vela molecular
ridge, Yamaguchi et al.\ (\cite{yam99}) found that the average luminosities
of IRAS sources in clouds associated with \HII\ regions and in clouds far
from \HII\ regions are $780~L_{\odot}$ and $63~L_{\odot}$ respectively.

Several processes, presented in Sect.~\ref{possproc}, have been proposed for
triggering  star formation at the peripheries of \HII\ regions 
(Elmegreen~\cite{elm98}). We are particularly interested in the `collect and collapse'
process because {\em it permits the formation of massive objects -- stars or
clusters}. This process, first proposed by Elmegreen \& Lada (\cite{ elm77})
has not, until now, been convincingly confirmed. The aim of the present paper is to
propose a list of carefully selected candidate regions which are likely to
be examples of this process at work. We present our selection criteria in
Sect.~\ref{selection}, we discuss the general mid-IR features of the selected
regions in Sect.~\ref{midir}, and we briefly comment on the individual
regions in Sect.~\ref{comments}. We have observed most of these regions in
the near-IR and a few at millimetre wavelengths; these studies will be
presented in separate papers (one of which -- Deharveng et al.~\cite{deh03b}
-- has already been published), in which we will investigate 
the reality of the collect and collapse process.
Finally, we present a discussion and conclusion in Sect.~\ref{discconc}.

\section{Possible processes triggering star formation at the periphery
of an \HII\ region\label{possproc}}

Consider a massive first-generation star that forms an \HII\ region.
Due to the high pressure of the warm ionized gas relative to that of
the surrounding cold neutral material, the \HII\ region expands; its
expansion velocity is of the order of 11~km~s$^{-1}$ just after the
ionization of the gas and the formation of the initial Strömgren
sphere, and it decreases with time (Dyson \& Williams \cite{dys97}).

During the expansion of the \HII\ region various events may occur.
\begin{enumerate}
\item The \HII\ region may expand into an inhomogeneous medium
containing pre-existing molecular clumps. The pressure exerted by the
ionized gas on the surface of a clump can lead to its implosion, and to
the formation of a `cometary globule' surrounded by dense ionized gas
forming a `bright rim' (see Bertoldi \cite{ber89}, 
Bertholdi \& McKee \cite{ber90}, and
Lefloch \& Lazareff \cite{lef94} for a simulation of the evolution of
such a globule). During the short collapse phase, a shock front
progresses into the clump, leading to the formation of a dense core.
This collapse is followed by a transient phase of re-expansion and then
by a quasi-stationary cometary phase. During this last phase the
cometary globule has a dense head and a tail extending away from the
ionizing source; the globule slowly evaporates as the dense
ionized gas flows away from it. All the signs of recent star
formation are observed in the direction of cometary globules: IRAS
sources with protostellar-object colours, MSX point sources, near-IR
reddened objects, CO outflows, Herbig-Haro objects, \Halpha\ emission
stars, etc.\ (e.g.\ Lefloch et al.~\cite{lef97}, 
Sugitani et al.~\cite{sug89}, Ogura et al.~\cite{ogu01} and Thompson et al.~\cite{tho04}).
However, no model explains {\em where} star formation takes place (in the
core, or at its periphery) or {\em when} (during the maximum compression 
phase, or earlier).
\item The ionization front of the \HII\ region is supersonic and is
preceded by a shock front in the neutral gas. With time, neutral
material accumulates between these two fronts. This compressed, shocked
material may become dynamically unstable on a {\em short} internal-crossing
timescale. This configuration was simulated by García-Segura \&
Franco (\cite{gar96}): the shell of neutral material fragments,
creating clumps which are afterward slowly eroded by the ionization
front, leading to the formation of cometary globules and bright rims.
The difference with the previous process is that the dense clumps do
not pre-exist, but are formed as the result of instabilities in
the collected layer.
\item The compressed shocked layer may become gravitationally unstable
along its surface, on a {\em long} timescale. This is the `collect and
collapse' model, first proposed by Elmegreen \& Lada (\cite{elm77}).
This process allows the formation of massive fragments. For example,
according to Whitworth et al.\ (\cite{whi94}), the compressed layer
around the \HII\ region formed by an O7 star, evolving in a medium of
10$^3$~cm$^{-3}$, with a sound speed of 0.5~km~s$^{-1}$ in the layer,
will become unstable after about 3 Myr; about seven fragments will
form, each with a mass $\sim600~M_{\odot}$. These fragments are dense
and they will very quickly fragment in turn, leading to the formation
of a cluster of coeval stars.

If the collected layer is not destroyed quickly by dynamical
instabilities, a large quantity of material accumulates within it. This
is the reason why its fragmentation (as a consequence of gravitational
instabilities) produces {\it massive} fragments -- massive enough to
form massive stars and/or clusters. Also, these fragments are {\em
regularly spaced around the H\,{\scriptsize{II}} region}.
\item The \HII\ region may form and evolve within a supersonic
turbulent molecular medium (Elmegreen et al.~\cite{elm95}). Clumps are
formed by turbulence compression in the gas. As the ionization front
and its associated shock move into the cloud, these clumps are
collected into the compressed layer; they merge into a few massive
postshock cores. These may collapse gravitationally, forming a cluster.
In this case, the dense cores would be {\em randomly} distributed
around the \HII\ region.
\item A completely different process is proposed by Fukuda \& Hanawa 
(\cite{fuk00}) who consider the interaction between an \HII\ region and
a nearby filamentary molecular cloud. As the \HII\ region expands, the
cloud is pinched and separated into two parts. A gravitational
instability induces the formation of two cores along the filament axis.
As this process is recursive, core formation -- and, hence, possibly
star formation -- propagates along the filament axis. 
\end{enumerate}

\section{Selection of candidate \HII\ regions \label{selection}}
 
To show that the collect and collapse mechanism is at work in a given 
star-formation zone, the
following predictions must be verified by observations: 
\begin{itemize}
   \item A neutral compressed layer should surround the \HII\ region.
   Being quite dense, the material must be molecular; and being a thin
   spherical shell, the layer should appear as a ring when projected on
   the plane of the sky. Hence {\em the compressed layer should be
   observed as a ring of molecular line emission at millimetre
   wavelengths}. Furthermore, since the layer contains dust, {\em it
   should also be observed as a ring of mid-IR and millimetre
   emissions}.
   \item The fragments should be dense, massive and {\em regularly
   spaced along the surface of the shell}.
   \item  Second generation stars or clusters, formed in the expanding
   layer, will have retained the velocity of the material in which they
   formed. Thus {\em they should be observed in the direction of the
   layer}, or slightly in front of it (on the neutral side) if the
   layer is slowing down.
\end{itemize}

The presence of several fragments {\em regularly} spaced along the
compressed layer is a strong argument in favour of the collect and
collapse process, as it allows us to reject processes involving 
pre-existing molecular clumps, or clumps formed by random turbulence.

None of the examples proposed by Elmegreen (\cite{elm98}) to illustrate
the collect and collapse model are completely convincing, as the
morphologies of these regions are complex, so that it is impossible
to verify if the above conditions are satisfied. If we want to
prove that the collect and collapse process works, we need to study
objects with a very simple morphology, where a clearly defined ionization
front separates the ionized gas from the dense neutral surrounding
medium. Therefore we have used the following criteria to select such
candidate regions:
\begin{itemize}
  \item A nearly spherical \HII\ region around an exciting star
  or cluster.
  \item A dust-emission ring surrounding the ionized gas. We have used
  the MSX Band A survey at 8.3~$\mu$m (see Sect.~\ref{midir}) to search
  for such rings. As explained in Sect.~\ref{midir}, this emission
  comes mainly from large molecules -- polycyclic aromatic hydrocarbons
  (PAHs) -- situated in the photo-dissociation region surrounding the
  ionized gas. The presence of this dust ring indicates that dense
  neutral gas surrounds the \HII\ region.
  \item An MSX point source in the direction of the dust ring. (We call
  a `point source' any source listed in the MSX Point Source Catalog 
  (Egan et al.~\cite{ega99}, \cite{ega03}), even though some are in fact resolved by
  MSX.) We expect these MSX point sources to trace small dust grains
  that are locally heated by an embedded cluster or by a massive
  object.

  The dense clumps (and subsequently the clusters) 
  resulting from the fragmentation of the compressed layer 
  can be located anywhere in the spherical shell. We realize that by
  rejecting those regions with MSX point sources which happen to
  project {\em inside} the dust ring, we are eliminating many legitimate
  cases where sources happen to be located more nearly in front of or
  behind the \HII\ region. If `nearly' means, say, within $45\degr$ of
  the line of sight, then 70.7\% of the shell projects on the ring. 
  If there are three point sources, the chances are only
  $0.707^3$, or 35\%, that none falls inside the ring. This is a high
  price to pay, but it nearly guarantees that a clump or cluster really
  lies in the vicinity of the compressed layer.
  \item Red stars or clusters associated with the 
  MSX point sources. We have used the 2MASS Survey to search for these
  objects.
\end{itemize}

For northern hemisphere regions, we have used the NVSS radio 
continuum survey at 1.4~GHz (Condon et al.~\cite{con98}), 
to see if ultracompact (UC) radio sources are present in
the direction of the MSX point sources.

Table~1 lists the selected sources. Column 1 gives the name of the 
\HII\ region surrounded by the MSX Band A emission ring, and columns 2 
and 3 the (approximate) coordinates of its centre. Columns 4 and 5 give the
coordinates of the MSX point source(s) observed in the direction of the
ring (sometimes several are observed; we are interested here by the
brighter ones). Column 6 gives the corresponding IRAS source, and
column 7 the distance of the region. Column 8 indicates the
observations which we have subsequently made of these regions.

The \HII\ region Sh~104 is the prototype of such objects (see its
description in Sect.~\ref{comments}). In the first of our planned
detailed studies, already published (Deharveng et al.\ \cite{deh03b}),
we have shown that this region is a good illustration of the collect
and collapse process. Our present aim is to find other, similar
regions susceptible to confirm this process and provide observations 
to further constrain the model. 

\begin{table*}
\caption{Selected regions.}
\begin{tabular}{rll@{}l@{}ll@{}l@{}ll@{}l@{}ll@{}l@{}lccl}
 \hline\hline
\multicolumn{8}{c}{Central \HII\ region} & \multicolumn{6}{c}{MSX point source} & IRAS & Distance & Comments  \\
  \hline
 1 & Sh~104         &    20$^{\rm h}$&17$^{\rm m}$&42\frs0 &   +36\degr&45\arcmin&25\arcsec&    20$^{\rm h}$&17$^{\rm m}$&56\frs6 &   +36\degr&45\arcmin&39\arcsec                 & 20160+3636   & 4.0      & {\footnotesize 1,\ 3,\ 4} \\
 2 & X              &    20          &20          &48.05  &   +39     &34       &43        &    20          &20          &46.1   &   +39     &35                &14                &              & ?        & {\footnotesize 1}         \\
 3 & Y              & \ \ 1          &01          &09.5   &   +62     &54       &08        & \ \ 1          &01          &08.4   &   +62     &54                &44                & 00580+6238   & 1.9      & {\footnotesize 1}         \\
 4 & Sh~212         & \ \ 4          &40          &38.9   &   +50     &27       &44        & \ \ 4          &40          &27.2   &   +50     &28                &29                & 04366+5022   & 7.1      & {\footnotesize 1}         \\
 5 & Sh~217         & \ \ 4          &58          &45.4   &   +47     &59       &58        & \ \ 4          &58          &30.3   &   +47     &58                &33$^{\mathrm{a}}$ & 04547+4753   & 5.0      & {\footnotesize 3,\ 4}     \\
 6 & Sh~219         & \ \ 4          &56          &10.5   &   +47     &23       &36        & \ \ 4          &56          &03.3   &   +47     &22                &57                & 04523+4718   & 5.0      & {\footnotesize 3}         \\
 7 & Z near Sh~230  & \ \ 5          &23          &09.95  &   +33     &58       &13        & \ \ 5          &23          &03.7   &   +33     &58                &31                & 05197+3355   & 1.8--3.2 & {\footnotesize 1,\ 4}     \\
 8 & Sh~241         & \ \ 6          &03          &58.6   &   +30     &15       &25        & \ \ 6          &03          &54.0   &   +30     &14                &49                & 06006+3015   & 4.7      & {\footnotesize 1,\ 4}     \\
 9 & Sh~259         & \ \ 6          &11          &27.7   &   +17     &26       &17        & \ \ 6          &11          &23.7   &   +17     &26                &29                & 06084+1727   & 8.3      & {\footnotesize 1,\ 4}     \\
10 & RCW~34         & \ \ 8          &56          &28.0   & $-$43     &06       &00        & \ \ 8          &56          &27.7   & $-$43     &05                &47                & 08546$-$4254 & 3.1      & {\footnotesize 2}         \\
11 & RCW~40         & \ \ 9          &02          &22.0   & $-$48     &41       &55        & \ \ 9          &02          &07.1   & $-$48     &43                &27                &              & 2.4      & {\footnotesize 5}         \\
12 & RCW~40         &                &            &       &           &         &          & \ \ 9          &02          &28.5   & $-$48     &39                &14                & 09007$-$4827 &          & {\footnotesize 5}         \\
13 & Dutra~45       &    10          &19          &51.4   & $-$58     &04       &09        &    10          &20          &15.7   & $-$58     &03                &57                & 10184$-$5748 & 4.5      & {\footnotesize 2}         \\
14 & Dutra~46       &    10          &29          &31.0   & $-$57     &26       &45        &    10          &29          &32.4   & $-$57     &26                &36                & 10276$-$5711 & 6.2      & {\footnotesize 2}         \\
15 & RCW~71         &    12          &50          &21.1   & $-$61     &34       &58        &    12          &50          &25.0   & $-$61     &35                &34                & 12474$-$6119 & 2.1      & {\footnotesize 5}         \\
16 & RCW~79         &    13          &40          &17.0   & $-$61     &44       &00        &    13          &40          &53.1   & $-$61     &45                &51                & 13374$-$6130 & 4.0      & {\footnotesize 2,\ 5}     \\
17 & RCW~82         &    13          &59          &29.0   & $-$61     &23       &40        &    13          &59          &57.1   & $-$61     &24                &37                & 13563$-$6109 & 2.9      & {\footnotesize 2}         \\
18 & RCW~82         &                &            &       &           &         &          &    13          &59          &03.6   & $-$61     &22                &17                & 13555$-$6107 &          &                           \\
19 & RCW~120        &    17          &12          &23.1   & $-$38     &27       &43        &    17          &12          &40.9   & $-$38     &27                &08                & 17092$-$3823 & 1.3      & {\footnotesize 5}         \\
  \hline
   &               &             &            &       &           &         &         &             &            &       &           &         &                  &              &          &
\end{tabular}

$^{\mathrm{a}}$ Not in the Point Source Catalog; we have measured its position on the MSX Band A frame. \\

\noindent Comments\\
{\footnotesize 1. $JHK$ observations at CFHT.                       }   \\
{\footnotesize 2. $JHK_S$ observations with the NTT telescope at ESO.} \\
{\footnotesize 3. molecular line emission observations at IRAM.    } \\
{\footnotesize 4. 1.2~mm continuum emission at IRAM.               }  \\
{\footnotesize 5. 1.2~mm continuum emission with the SEST at ESO.  }

\end{table*}

\section{Mid-IR emission of the selected sources \label{midir}}

The mid-IR emission from Galactic \HII\ regions depends on their
evolutionary status (UC or more evolved) and has been studied by many
authors, both spectroscopically and with imaging, using ground-based
and space facilities. The main results of these studies can be
summarized as follows: the spectra of UC \HII\ regions show a continuum
emission increasing with wavelength, and strong silicate absorption
features centred at 9.7 and 18\,$\mu$m (cf.\ Faison et al.\
\cite{fai98}). More evolved \HII\ regions also show a continuum
emission increasing with wavelength, but superimposed on the continuum
are mid-IR bands, the unidentified infrared bands (UIBs), centred at
6.2, 7.7, 8.6, 11.3 and 12.7~$\mu$m. These emission bands have been
attributed to large molecules such as polycyclic aromatic hydrocarbons,
PAHs (Léger \& Puget \cite{leg84}). Inside the ionized gas the dominant
emission is the continuum emission from small grains, stronger for
earlier-type exciting stars; beyond the ionization front, the emission
is predominantly from large molecules such as PAHs (cf.\ Zavagno \&
Ducci \cite{zav01}). Small grains can survive in strong UV fields,
while large molecules cannot (cf.\ Crété et al.~\cite{cre99}, Peeters
et al.~\cite{pee02}). Atomic fine-structure lines and hydrogen
recombination lines are also present in the spectra of \HII\ regions (
see Peeters et al.~\cite{pee02} for UC \HII\ regions).

The mid-IR spectra of Herbig Ae/Be stars (intermediate mass 
pre-main-sequence stars) show large star-to-star differences. 
Most present a continuum emission increasing with wavelength; some
exhibit a strong silicate emission near 10~$\mu$m, others PAH emission
bands (Bouwman et al.\ \cite{bou01}, Meeus et al.\ \cite{mee01}).

Mid-IR emission associated with \HII\ regions
has been studied by Crowther \& Conti (\cite{cro03} -- for
UC regions) and by Ghosh \& Ojah (\cite{gho02}) and
Conti \& Crowther (\cite{con03} -- for more evolved ones) using the
Midcourse Space eXperiment (MSX) with the on-board instrument
SPIRIT III (Price et al.\ \cite{pri01}). This experiment has
surveyed the entire Galactic plane ($|b| < 5\deg$) in four mid-IR
bands -- A, C, D and E -- centred at 
8.3, 12.13, 14.65 and 21.34~$\mu$m respectively, with an
angular resolution of about 18$\arcsec$. The MSX Band A 
includes the dominant UIBs at 7.7 and
8.6\,$\mu$m. This band also includes the silicate 
feature at 9.7\,$\mu$m. The MSX Band C includes the 
UIBs at 11.3 and 12.7\,$\mu$m, as well as continuum emission. 
In UC \HII\ regions, the continuum 
dominates and increases continuously from Band C to Band E.

\begin{figure*}
     \includegraphics[width=170mm]{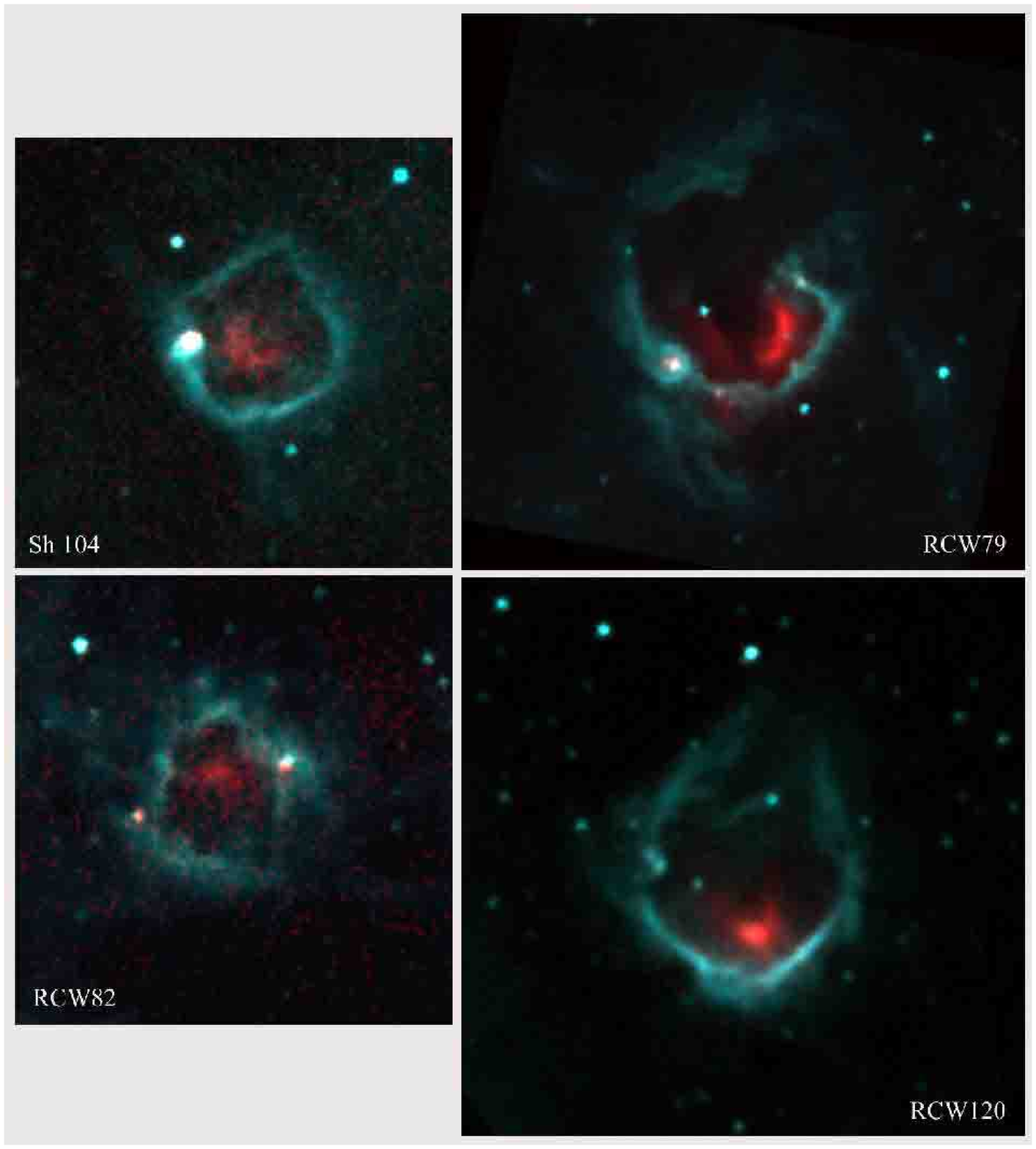}
     \caption{Colour composite MSX images of four regions. The
     Band A emission centred at 8.3~$\mu$m, dominated by the UIBs,
     appears in turquoise. The Band E emission centred at
     21.3~$\mu$m, dominated by the continuum
     emission of small grains, appears in red. North is up and
     east is left. The field size is $20\arcmin \times 20\arcmin$
     for Sh~104, $30\arcmin \times 30\arcmin$ for RCW~79,
     $18\farcm 3 \times 18\farcm 3$ for RCW~82, and
     $23\farcm3 \times 23\farcm 3$ for RCW~120.}
\end{figure*}

We have used the MSX Bands A and E in order to characterize the mid-IR
emission of the extended sources in Table~1. Fig.~1 presents colour
composite images of four regions -- Sh~104, RCW~79, RCW~82 and RCW~120. The
Band A emission is displayed in turquoise and the Band~E emission in
red. The Band A emission, dominated by the UIBs, forms a ring around the 
\HII\ region. Band E, dominated by the continuum emission from
small grains, peaks {\em inside} the turquoise ring, in the direction of
the ionized gas. These images show the importance of studying objects
with simple morphologies, in which emissions of various origin are
clearly separated in the plane of the sky. These regions unambiquiously
demonstrate that: i) the UIB carriers do not survive in the ionized gas,
but are present in the photo-dissociation region where they are excited
by the UV radiation leaking from the \HII\ region; ii) the small grains
responsible for the mid-IR emission are not destroyed in the ionized
gas, and compete with the gas to absorb the Lyman continuum photons
emitted by the exciting stars.

We have also used the MSX fluxes (given in Table~2) 
of the point source(s) observed towards
the dust ring, to tentatively determine their nature,
either UC \HII\ regions or intermediate mass Herbig Ae/Be stars. As
shown previously, each type of object possesses a
characteristic mid-IR spectrum depending on its nature and/or
evolutionary status. Mid-IR colour-colour diagrams can therefore be used
to determine the nature of the observed point sources. Egan et 
al.~(\cite{ega01}) and Lumsden et al.~(\cite{lum02}) have combined near-IR
and mid-IR colour-colour diagrams to identify different kinds of MSX
point sources. They have shown that all the young sources (Herbig Ae/Be
stars, massive young stellar objects and UC \HII\ regions) are well
separated from the bulk of normal stars, a consequence of the large
extinction due to the fact that the young objects are embedded in
optically-thick dust clouds. These authors have also shown that the UC
\HII\ regions and the Herbig Ae/Be stars occupy different positions in
the F$_{14}$/F$_{8}$ versus F$_{21}$/F$_{14}$ and the 
F$_{14}$/F$_{12}$ versus F$_{21}$/F$_{8}$ diagrams
(Fig.~2). Their explanation is that the Herbig Ae/Be stars are the least
embedded of the young sources, which is reflected in their having bluer
colours than the other objects. An alternative explanation is that, due
to the fact that the spectra near 10~$\mu$m of some Herbig Ae/Be stars
are dominated by the silicate emission feature (Meeus et
al.~\cite{mee01}), their Band A emission is enhanced.

UC \HII\ regions have been clearly detected as radio continuum 
sources in the direction of the MSX point sources
associated with Sh~104, Sh~217 and RCW~79. However, few of the
other sources in Table~1 have been observed in the radio continuum
with an angular resolution high enough to allow the detection of
a UC \HII\ region -- if any -- at the border of the central \HII\ region.
Therefore we have used the mid-IR colours of the MSX point sources
to get an idea about their nature -- do they contain stars massive
enough to ionize the gas, or intermediate-mass stars? Fig.~2
shows the position of the MSX point sources of Table~2 in Lumsden
et al.'s~(\cite{lum02}) colour-colour diagrams. The two diagrams
give consistent information: some of the MSX point sources -- those
observed at the borders of Sh~212, Sh~259, RCW~34, Dutra~45 and RCW~82
(18) -- are probably associated with UC \HII\ regions.

We can also use the far-IR luminosities of the MSX point sources,
given in Table~2, to get additional information  
about the presence of massive stars (assuming that the
radiation of the young objects is reprocessed by dust, and emerges
in the far IR). The far-IR flux can be obtained from the observed
IRAS fluxes (Table~2) using the relation (Emerson~\cite{eme88} cited by
Lumsden et al.~\cite{lum02})

$F_{\rm FIR}=(20.6 F_{12}+7.54 F_{25}+4.58 F_{60} +1.76 F_{100})$~W~m$^{-2}$.

The far-IR luminosities are obtained from these fluxes using the
distances given in Table~1. Using the effective temperatures of
main-sequence stars and their radii, according to Smith et
al.~(\cite{smi02}), the luminosity  of a B1.5V
star is $17800~L_{\odot}$. No figures are given by these authors for a B2V
star. We will assume that any cluster containing a star massive
enough to form an \HII\ region (at least a B2V star) should be
more luminous than $\sim10000~L_{\odot}$. The far-IR luminosities
confirm that the sources detected at the border of Sh~212, Sh~259,
RCW~34 and Dutra~45 most probably contain stars massive enough to
ionize the gas. This is also probably the case for the sources
associated with Sh~241 and Dutra~46. The results are contradictory for RCW~40
(12) and RCW~82 (18): either they do not have the
colours of UC \HII\ regions or they are not luminous enough. 
And finally, both the MSX colours and the far-IR luminosities indicate that the 
point sources associated with RCW~82 (17) and RCW~120 do not 
contain massive stars.

\begin{table*}
\caption{IR luminosities and colours of the selected sources.}
\begin{tabular}{rll@{}l@{}l@{}lll@{}l@{}l@{}lr}
\hline\hline
 && \multicolumn{4}{c}{MSX point source} & \multicolumn{6}{c}{IRAS point source} \\
&   & \ \ \ \ $F_{8}$ & \ \ \ \ $F_{12}$ & \ \ \ \ $F_{14}$ & \ \ \ \ $F_{21}$& & \ \ \ \ $F_{12}$ & \ \ \ \ $F_{25}$
& \ \ \ \ $F_{60}$& \ \ \ \ $F_{100}$& $L_{\rm FIR}$\ \  \\
&  & \ \ \ \  (Jy) & \ \ \ \  (Jy) & \ \ \ \  (Jy) & \ \ \ \  (Jy) & & \ \ \ \  (Jy) & \ \ \ \  (Jy) & \ \ \ \  (Jy) & \ \ \ \  (Jy) & ($L_{\odot}$)\ \  \\
  \hline
 1 & Sh~104   &\ \ \ \ 8.39 &\ \ \ \ 10.45    &\ \ \ \ \ \ 7.36 &\ \ \ \ \ \ 41.54    & $20160+3636$ &\ \ \ \ 15.2\ \    &\ \ \ \    \ \ 94.2\ \       &\ \ \ \ \ \ 647      &\ \ \ \    1160 &  30000  \\
 2 & X        &\ \ \ \ 2.99 &\ \ \ \ \ \ 4.10 &\ \ \ \ \ \ 1.88 &\ \ \ \ \ \ \ \ 2.85 & $          $ &\ \ \ \            &\ \ \ \                      &\ \ \ \ \ \ \ \      &\ \ \ \ \ \     &         \\
 3 & Y        &\ \ \ \ 0.65 &\ \ \ \ \ \ 0.93 &\ \ \ \          &\ \ \ \ \ \ \ \      & $00580+6238$ &\ \ \ \ \ \ 1.91   &\ \ \ \ \ \ \ \ 1.76         &\ \ \ \ \ \ \ \ 28.5 &\ \ \ \ \ \ 108 &    400  \\
 4 & Sh~212   &\ \ \ \ 2.27 &\ \ \ \ \ \ 2.91 &\ \ \ \ \ \ 4.94 &\ \ \ \ \ \ 35.34    & $04366+5022$ &\ \ \ \ \ \ 2.87   &\ \ \ \    \ \ 54.2\ \       &\ \ \ \ \ \ 199      &\ \ \ \ \ \ 470 &  35000  \\
 5 & Sh~217   &\ \ \ \      &\ \ \ \ \ \      &\ \ \ \          &\ \ \ \ \ \ \ \      & $04547+4753$ &\ \ \ \ 10.4\ \    &\ \ \ \    \ \ 82.0\ \       &\ \ \ \ \ \ \ \ 36.0 &\ \ \ \ \ \ 367 &  13000  \\
 6 & Sh~219   &\ \ \ \ 1.60 &\ \ \ \ \ \ 2.11 &\ \ \ \ \ \ 2.01 &\ \ \ \ \ \ \ \ 3.00 & $04523+4718$ &\ \ \ \ \ \ 3.30   &\ \ \ \ \ \ \ \ 5.21         &\ \ \ \ \ \ \ \ 81.7 &\ \ \ \ \ \ 305 &   8000  \\
 7 & Z        &\ \ \ \ 2.08 &\ \ \ \ \ \ 2.66 &\ \ \ \ \ \ 1.35 &\ \ \ \ \ \ \ \ 3.58 & $05197+3355$ &\ \ \ \ \ \ 6.84   &\ \ \ \    \ \ 15.6\ \       &\ \ \ \ \ \ 270      &\ \ \ \ \ \ 581 &  2500-8000  \\
 8 & Sh~241   &\ \ \ \      &\ \ \ \ \ \      &\ \ \ \          &\ \ \ \              & $06006+3015$ &\ \ \ \ \ \ 1.94   &\ \ \ \    \ \ 11.8\ \       &\ \ \ \ \ \ 189      &\ \ \ \ \ \ 552 &  14000  \\
 9 & Sh~259   &\ \ \ \ 1.00 &\ \ \ \ \ \ 2.05 &\ \ \ \ \ \ 3.25 &\ \ \ \ \ \ 13.85    & $06084+1727$ &\ \ \ \ \ \ 3.22   &\ \ \ \    \ \ 24.8\ \       &\ \ \ \ \ \ 124      &\ \ \ \ \ \ 263 &  28000  \\
10 & RCW~34   &\ \ \ \ 7.47 &\ \ \ \ 17.37    &\ \ \ \ 20.64    &\ \ \ \ \ \ 75.97    & $08546-4254$ &\ \ \ \ \ \ 2.93   &\ \ \ \       194\ \ \ \ \ \ &\ \ \ \ 1340         &\ \ \ \    2100 &  34000  \\
11 & RCW~40   &\ \ \ \ 3.23 &\ \ \ \ \ \ 5.50 &\ \ \ \ \ \ 2.53 &\ \ \ \ \ \ \ \ 5.69 & $          $ &\ \ \ \            &\ \ \ \                      &\ \ \ \ \ \ \ \      &\ \ \ \ \ \     &         \\
12 & RCW~40   &\ \ \ \ 2.62 &\ \ \ \ \ \ 3.21 &\ \ \ \ \ \ 1.09 &\ \ \ \ \ \ \ \ 2.43 & $09007-4827$ &\ \ \ \ 16.0\ \    &\ \ \ \    \ \ 19.4\ \       &\ \ \ \ \ \ 995      &\ \ \ \    2620 &  17000  \\
13 & Dutra~45 &\ \ \ \ 6.84 &\ \ \ \ 16.96    &\ \ \ \ 36.49    &\ \ \ \ 197.2\ \     & $10184-5748$ &\ \ \ \ 31.8\ \    &\ \ \ \       349\ \ \ \ \ \ &\ \ \ \ 3330         &\ \ \ \    5410 & 178000  \\
14 & Dutra~46 &\ \ \ \ 2.71 &\ \ \ \ \ \ 5.49 &\ \ \ \ \ \ 4.62 &\ \ \ \ \ \ 10.33    & $10276-5711$ &\ \ \ \ 12.5\ \    &\ \ \ \    \ \ 61.4\ \       &\ \ \ \ \ \ 492      &\ \ \ \    1250 &  62000  \\
15 & RCW~71   &\ \ \ \ 2.28 &\ \ \ \ \ \ 4.77 &\ \ \ \ \ \ 2.10 &\ \ \ \ \ \ \ \ 4.36 & $12474-6119$ &\ \ \ \ \ \ 8.39   &\ \ \ \    \ \ 30.0\ \       &\ \ \ \ \ \ 548      &\ \ \ \    1420 &   7500  \\
16 & RCW~79   &\ \ \ \ 3.64 &\ \ \ \ 10.04    &\ \ \ \ 13.33    &\ \ \ \ \ \ 33.10    & $13374-6130$ &\ \ \ \ 25.1\ \    &\ \ \ \    \ \ 95.5\ \       &\ \ \ \ 1020         &\ \ \ \    2850 &  55000  \\
17 & RCW~82   &\ \ \ \ 1.34 &\ \ \ \ \ \ 1.98 &\ \ \ \ \ \ 1.99 &\ \ \ \ \ \ \ \ 5.99 & $13563-6109$ &\ \ \ \ \ \ 2.76   &\ \ \ \    \ \ 12.60         &\ \ \ \ \ \ 104      &\ \ \ \ \ \ 319 &   3000  \\
18 & RCW~82   &\ \ \ \ 0.64 &\ \ \ \ \ \ 2.44 &\ \ \ \ \ \ 5.48 &\ \ \ \ \ \ \ \ 6.88 & $13555-6107$ &\ \ \ \ \ \ 7.0\ \ &\ \ \ \    \ \ 14.7\ \       &\ \ \ \ \ \ 136      &\ \ \ \ \ \ 576 &   5000  \\
19 & RCW~120  &\ \ \ \ 2.76 &\ \ \ \ \ \ 3.42 &\ \ \ \ \ \ 2.08 &\ \ \ \ \ \ \ \ 2.92 & $17092-3823$ &\ \ \ \ \ \ 4.66   &\ \ \ \ \ \ \ \ 8.43         &\ \ \ \ \ \ \ \ 73.1 &\ \ \ \    3920 &   4000  \\
  \hline
\end{tabular}
\end{table*}


\begin{figure*}
    \includegraphics[width=85mm]{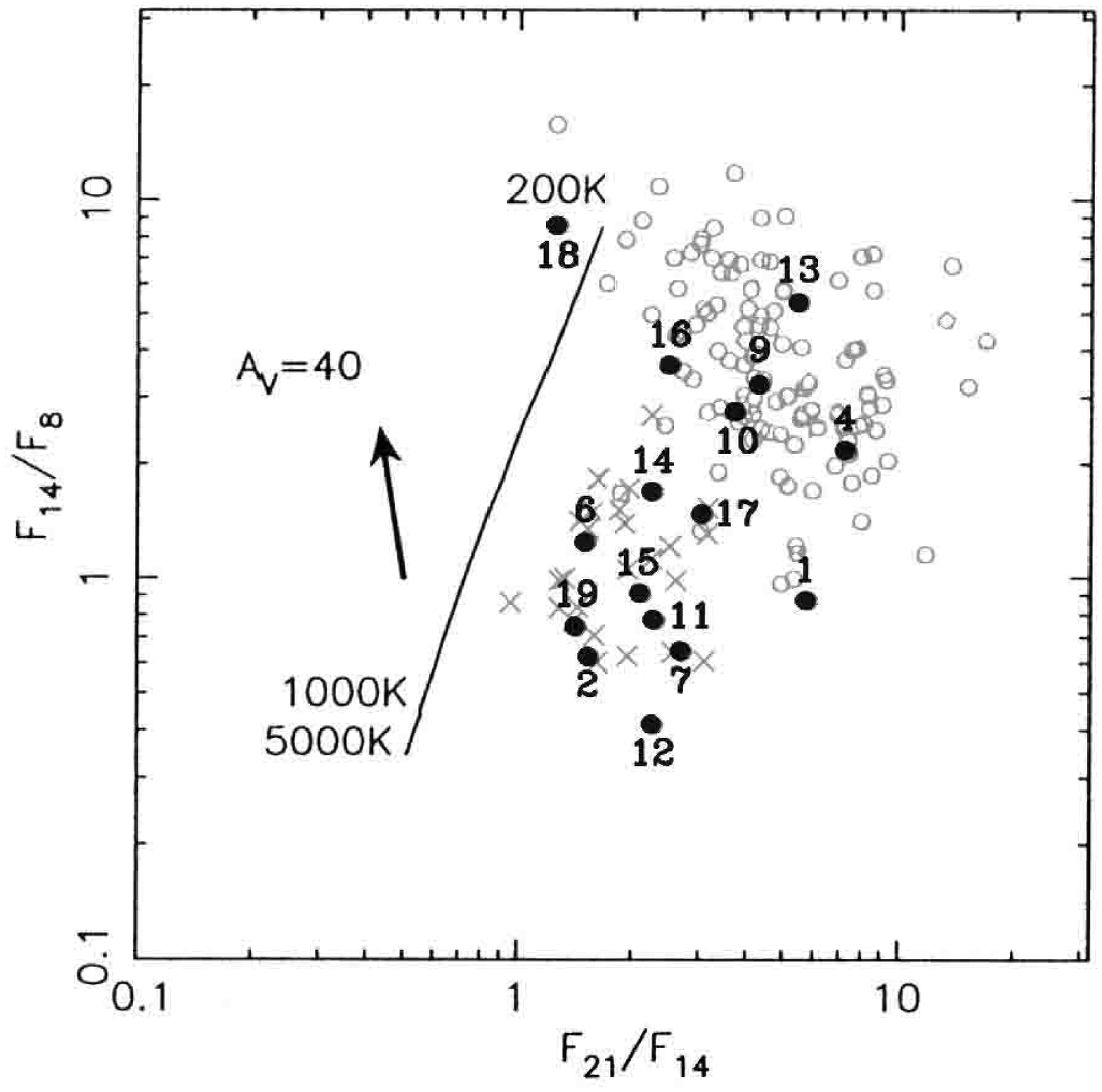}
    \includegraphics[width=85mm]{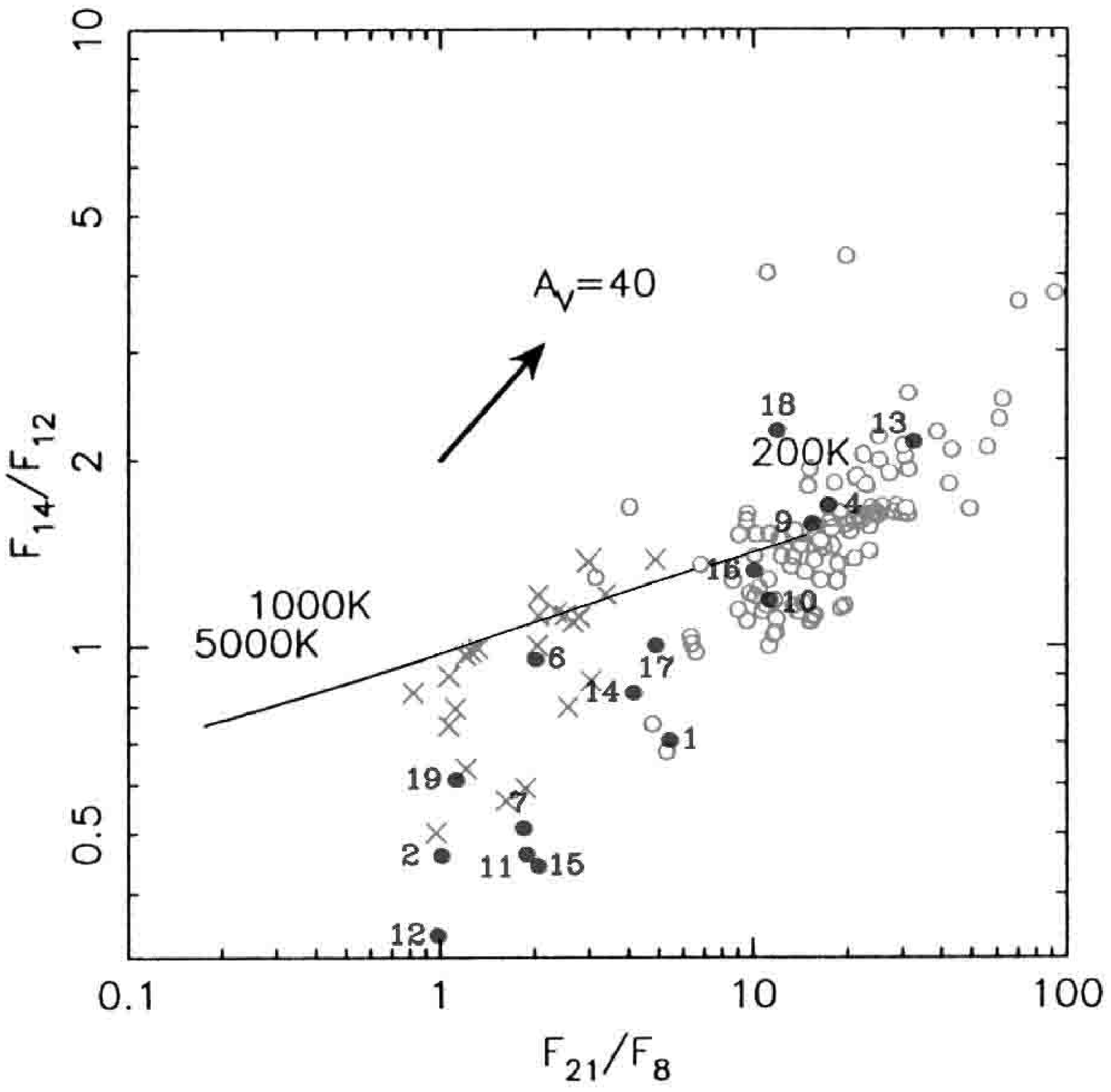}
    \caption{Mid-IR colour-colour plots of the MSX point sources (black
    filled circles); the sources are identified by their numbers in
    Tables~1 and 2. The underlying figure is from Lumsden et al.~(\cite{
    lum02}), with their original symbols changed to grey; the grey
    crosses are Herbig Ae/Be stars, and the empty grey circles are
    UC \HII\ regions. (Other objects in the original figure but
    not discussed here have been removed.)}
\end{figure*}

\section{Comments on individual sources\label{comments}}

Figs 3 to 19 present, for almost all the regions of Table~1, the MSX
Band A emission, as contours superimposed on an optical image of the
\HII\ region (extracted from the DSS-2red survey or from the SuperCOSMOS
H-alpha survey, Parker \& Phillipps \cite{par98}). Exceptions are 
source~X and Dutra~45, with no optically visible central \HII\ region; in these
cases we have used the radio continuum emission to locate the ionized
gas. A few comments are given for each source, mainly about the
morphologies of the various components of the complexes. 

\subsection{Northern hemisphere regions}

{\bf Sh~104} (Deharveng et al.~\cite{deh03b}) is the prototype of the
sort of \HII\ region we are looking for in order to illustrate and test the
collect and collapse model. Sh~104 is a 7$\arcmin$ diameter optically
visible \HII\ region, excited by an O6V central star. This region is
spectacular in the MSX Band A survey (Fig.~1), where a complete ring of
dust emission surrounds the ionized gas. An MSX point source lies
exactly on the dust ring. Near-IR observations show that this source
harbours a deeply embedded cluster which, since it is associated with a
UC \HII\ region, must contain at least one massive OB2 star. Our IRAM
observations show that a ring of molecular material surrounds Sh~104.
This ring contains four dense fragments {\em regularly spaced along the
ring} and each formed of several dense cores. The cluster is embedded
within the brightest fragment, whose mass is $\sim670~\msol$.

\begin{figure}
  \includegraphics[width=85mm]{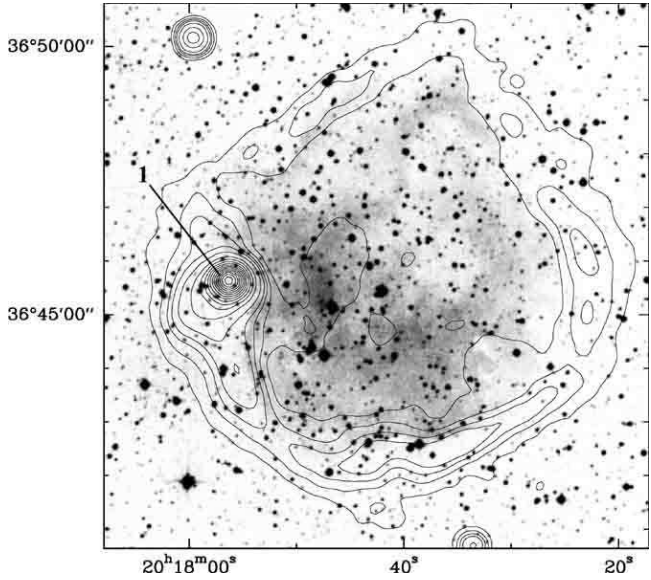}
  \caption{Sh~104. MSX Band A emission (contours) superimposed on
  a DSS-2 red image. The first contour levels are at $3\times10^{-6}$   
  and $4\times10^{-6}$, and then 
  increase in  steps of $4\times10^{-6}$ W m$^{-2}$ sr$^{-1}$.}
\end{figure}

\begin{figure}
  \includegraphics[angle=270,width=85mm]{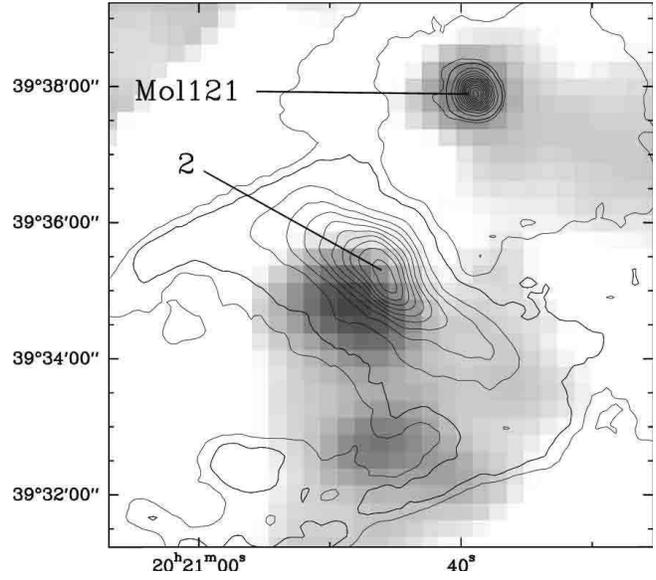}
  \caption{Source X. MSX Band A contours superimposed on a 1.4~GHz continuum. 
  The first contour levels are at $4\times10^{-6}$   
  and $5\times10^{-6}$, and then 
  increase in  steps of $3\times10^{-6}$ W m$^{-2}$ sr$^{-1}$.}
\end{figure}

{\bf Source X} The central \HII\ region is not optically visible, but is
a radio continuum source of flux density 37.3~mJy at 1.4~GHz (NVSS
Survey). Its distance is unknown. MSX Band A emission surrounds this
\HII\ region on its north, east and south sides. The brightness of the
ring is clearly enhanced in the north-east; a cluster of near-IR objects
lies in this direction. Another bright and very red MSX point source,
observed $3\farcm2$ north of the ring, corresponds to Mol~121 (Molinari
et al.~\cite{mol98}), a UC \HII\ region containing a small cluster of
near-IR objects (as shown by our CFHT $JHK$ images).

{\bf Source Y} is a faint unnamed optical nebulosity of diameter
$\sim1\farcm0$. It corresponds to a very faint radio continuum source of
flux density 3.5~mJy at 1.4~GHz (NVSS Survey). The distance of this
region is 1.9~kpc (Kerton~\cite{ker02}). A half-ring of MSX Band A
emission is observed exterior to the \HII\ region. This emission is
enhanced just north of the source, and several bright near-IR stars are
observed in this direction (but no emission is detected at 14~$\mu$m and
21~$\mu$m). A molecular cloud, of 185~\msol, also lies just north of
source Y (Kerton~\cite{ker02}).

\begin{figure}
  \includegraphics[angle=270,width=85mm]{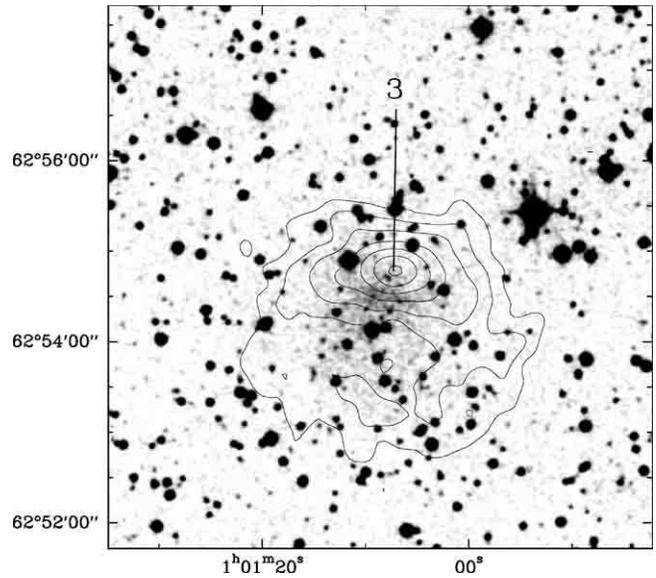}
  \caption{Source Y. MSX Band A contours superimposed on
  a DSS-2 red image. The first contour levels are at $1.4\times10^{-6}$  
  and $2\times10^{-6}$, and then 
  increase in  steps of $1\times10^{-6}$ W m$^{-2}$ sr$^{-1}$.}
\end{figure}

{\bf Sh~212} is a bright optical \HII\ region, of diameter
$\sim5\arcmin$, nearly circular around its central exciting cluster. It
is a high excitation region, ionized mainly by an O5.5neb star, at a
distance of 7.1~kpc (Caplan et al.\ \cite{cap00}). A ring of MSX
emission surrounds the bright northern part of the \HII\ region (but
not all of it). A bright MSX point source lies along this ring, 
north-west of Sh~212. An optical reflection nebula and a bright 
near-IR source are observed in this direction.

\begin{figure}
  \includegraphics[angle=270,width=85mm]{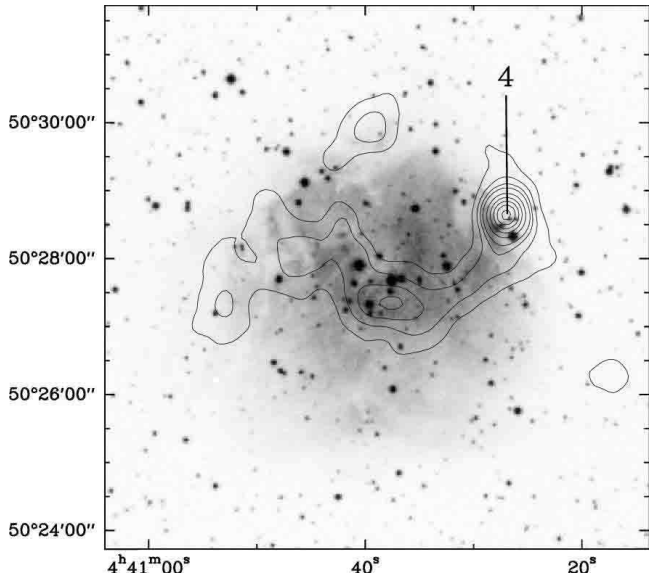}
  \caption{Sh~212. MSX Band A contours superimposed on
  a DSS-2 red image. The first contour levels are at $2\times10^{-6}$, 
  $3\times10^{-6}$, $4\times10^{-6}$ and $5\times10^{-6}$, and then 
  increase in  steps of $2\times10^{-6}$ W m$^{-2}$ sr$^{-1}$.}
\end{figure}

{\bf Sh~217} is an elliptical optical \HII\ region,
$10\arcmin \times 7\farcm5$ in size, excited by a central O9V star. Its
distance is 5.0~kpc. There is a half-ring of MSX Band A emission south
and west of this \HII\ region. A bright MSX point source is observed in
the middle of the arc. This corresponds to a deeply embedded near-IR
cluster which contains at least one OB2 star, as it is associated
with a UC \HII\ region (Deharveng et al.~\cite{deh03a}). Molecular
condensations are present south and west of this \HII\ region; the
brightest one lies in the direction of the cluster 
(Brand et al.~in preparation). Half a ring of
low density atomic hydrogen curves around the ionized gas to the north
and east (Roger \& Leahy~\cite{rog93}).

\begin{figure}
  \includegraphics[angle=270,width=85mm]{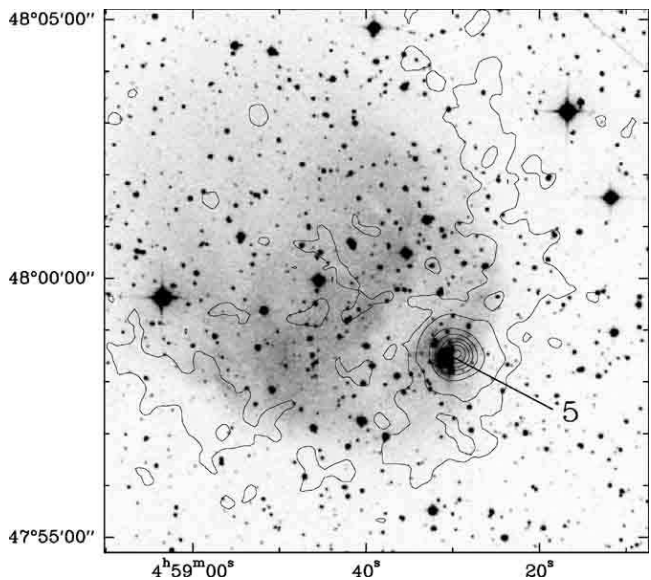}
  \caption{Sh~217. MSX Band A contours superimposed on
  a DSS-2 red image. The first contour levels are at $1\times10^{-6}$  
  and $2\times10^{-6}$, and then 
  increase in  steps of $5\times10^{-6}$ W m$^{-2}$ sr$^{-1}$.}
\end{figure}

{\bf Sh~219} is a spherical \HII\ region, of diameter
$1\farcm5$, excited by a B0V star. Its distance is 5.0~kpc. Half a ring
of MSX dust emission encloses the \HII\ region on its south and west
sides. This ring is enhanced in the middle of the arc, where a deeply
embedded cluster is observed elongated along the ring (Deharveng 
et al.~\cite{deh03a}). An \Halpha\ emission star with a near-IR excess, and
affected by a visual extinction $\sim24$~mag, lies near the
centre of the cluster. The presence nearby of a UC \HII\ region 
(Leahy~\cite{lea97}) remains to be confirmed. A molecular condensation is
observed in the direction of the cluster (Lefloch et al.~in
preparation). Elsewhere the ionized gas is surrounded by a low density
atomic hydrogen ring (Roger \& Leahy~\cite{rog93}).

\begin{figure}
  \includegraphics[angle=270,width=85mm]{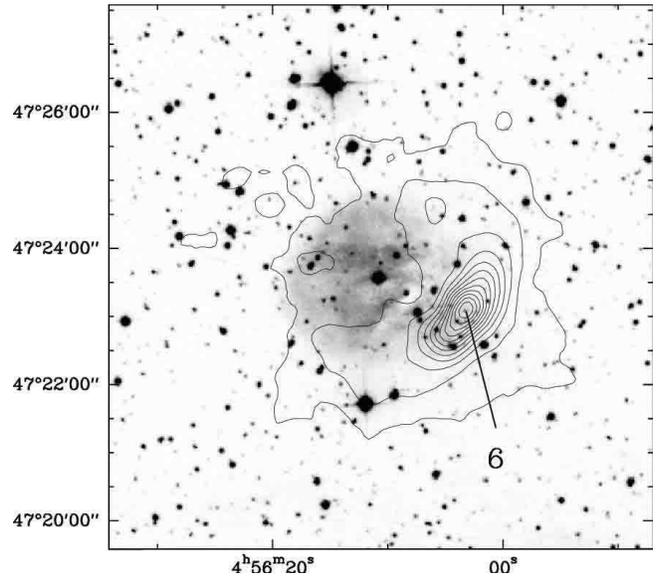}
  \caption{Sh~219. MSX Band A contours superimposed on
  a DSS-2 red image. The contour levels begin at $1\times10^{-6}$ and 
  increase in  steps of $1\times10^{-6}$ W m$^{-2}$ sr$^{-1}$.}
\end{figure}

\begin{figure}
  \includegraphics[angle=270,width=85mm]{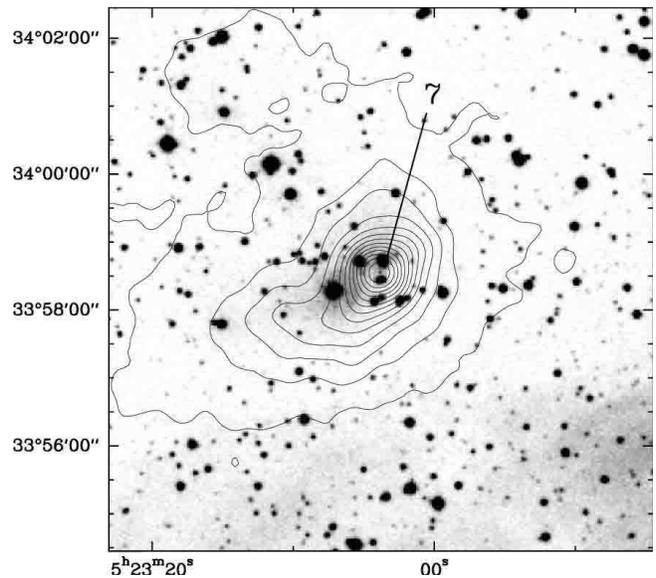}
  \caption{Source Z. MSX Band A contours superimposed on
  a DSS-2 red image. The contour levels begin at $1\times10^{-6}$ and 
  increase in  steps of $1.5\times10^{-6}$ W m$^{-2}$ sr$^{-1}$.}
\end{figure}

{\bf Source Z} is a small unnamed optical nebulosity, of diameter
$\sim1\farcm5$, located at the edge of the large diffuse \HII\ region
Sh~230. It is also a faint radio continuum source (Carpenter et al.\ \cite{
car90} and NVSS survey). The distance of this region is very uncertain:
1.8~kpc is its kinematic distance (Wouterloot \& Brand~\cite{wou89})
and 3.2~kpc is the photometric distance of the nearby \HII\ regions 
(with similar velocities) Sh~236 and Sh~237 (Carpenter et al.\ \cite{
car90}). Source Z is bordered on its western and southern sides by a
half-ring of dust emission. An MSX point source is observed on the ring,
west of source Z. A near-IR cluster lies in the same direction. A
molecular cloud of 870~\msol\ lies nearby, in the direction of the IRAS
source (Carpenter et al.~\cite{car90}, for a distance of 3.2~kpc).

\begin{figure}
  \includegraphics[angle=270,width=85mm]{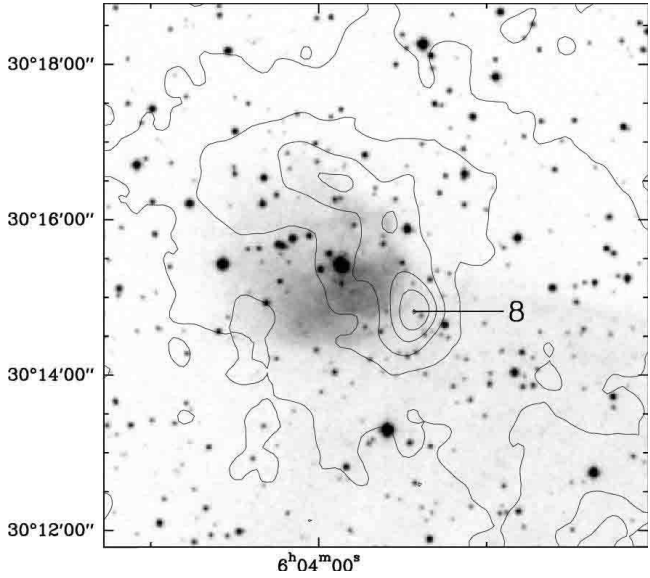}
  \caption{Sh~241. MSX Band A contours superimposed on
  a DSS-2 red image. The contour levels begin at $1\times10^{-6}$ and 
  increase in  steps of $1.5\times10^{-6}$ W m$^{-2}$ sr$^{-1}$.}
\end{figure}

\begin{figure}
  \includegraphics[angle=270,width=85mm]{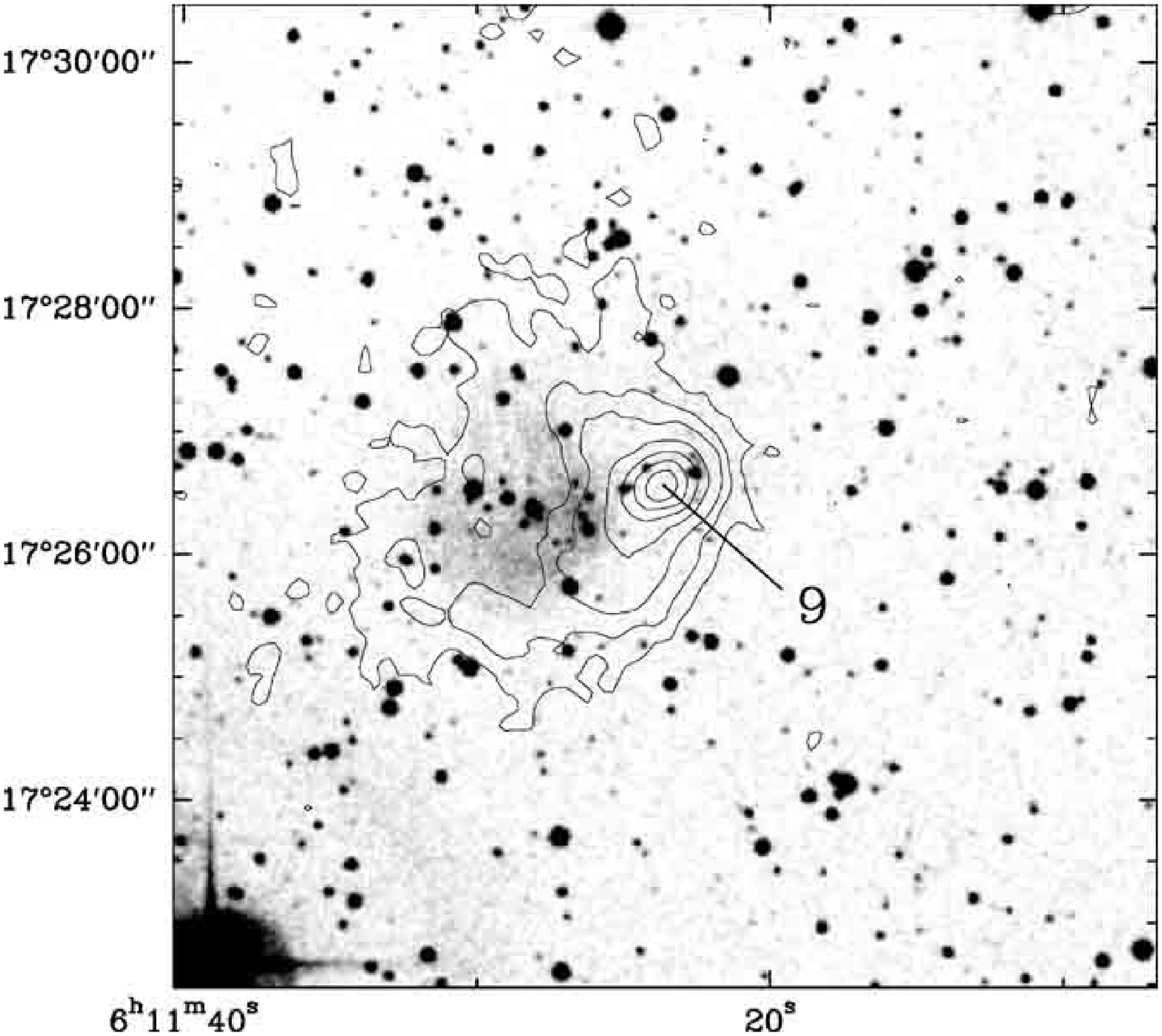}
  \caption{Sh~259. MSX Band A contours superimposed on
  a DSS-2 red image. The contour levels begin at $1.5\times10^{-6}$ and 
  increase in  steps of $1.5\times10^{-6}$ W m$^{-2}$ sr$^{-1}$.}
\end{figure}

\begin{figure}
  \includegraphics[angle=270,width=85mm]{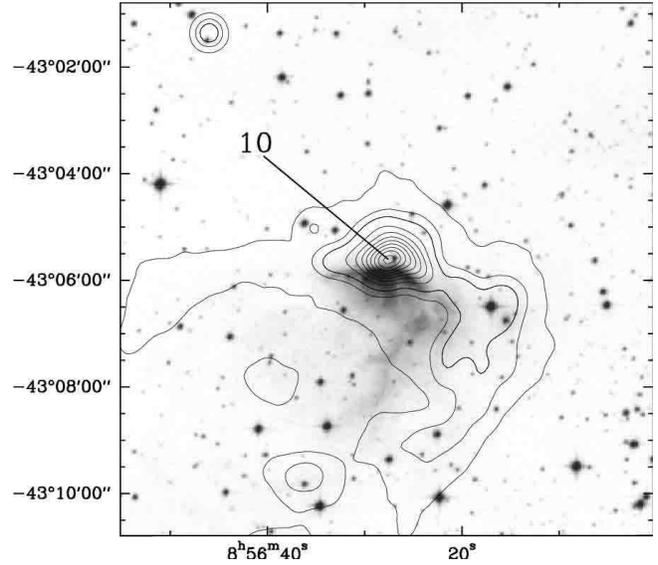}
  \caption{RCW~34. MSX Band A contours superimposed on
  a DSS-2 red image. The first contour levels are at $3\times10^{-6}$, 
  $6\times10^{-6}$ and $9\times10^{-6}$, and then 
  increase in  steps of $6\times10^{-6}$ W m$^{-2}$ sr$^{-1}$.}
\end{figure}

\begin{figure}
  \includegraphics[angle=270,width=85mm]{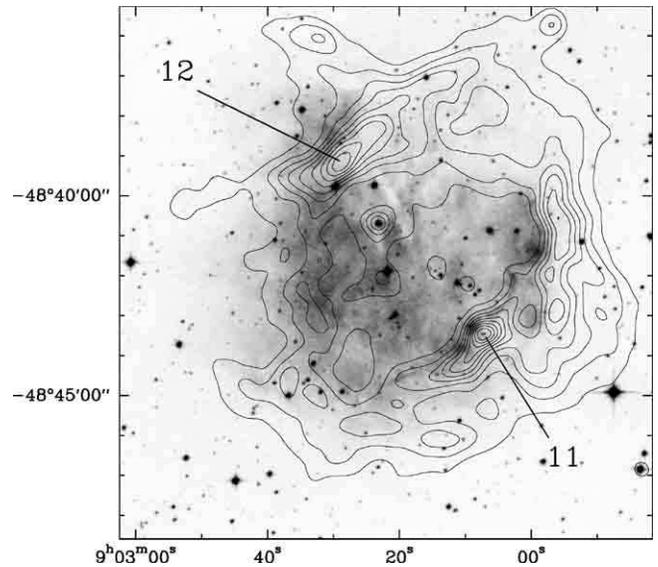}
  \caption{RCW~40. MSX Band A contours superimposed on
  a DSS-2 red image. The contour levels begin at $5\times10^{-6}$ and 
  increase in  steps of $3\times10^{-6}$ W m$^{-2}$ sr$^{-1}$.}
\end{figure}

{\bf Sh~241} is a low-brightness optical \HII\ region of diameter 
$10\arcmin$. A brighter (but unnamed), nearly spherical \HII\ region of
diameter $\sim2\arcmin$ lies at its north-east corner. We are interested
in this brighter region, excited by a central O9V star, and lying at a distance of 
4.7~kpc (Moffat et al.~\cite{mof79}). Half a ring of dust emission borders
this brighter \HII\ region to the north and east. An MSX point source lies at the
southern extremity of the arc (which is not listed in the MSX Point
Source Catalog), and a near-IR cluster is also observed in this
direction.  Nearby is a dense molecular core which has been
observed in the CS and HCN lines (Plume et al.\ \cite{plu92}, \cite{plu97}, 
Pirogov \cite{ pir99}), and mapped at 350~$\mu$m (Mueller et al.
\ \cite{mue02}). An H$_{2}$O maser (Cesaroni et al.~\cite{ces88},
Henning et al.~\cite{hen92}) and a molecular outflow 
(Wu et al.\ \cite{wu99}) have also been detected in the direction of 
the MSX and IRAS point sources, demonstrating that 
massive-star formation is presently taking place.

{\bf Sh~259} is a low-brightness optical \HII\ region, of diameter
$\sim2\arcmin$. It lies in the general direction of the Gemini OB1 molecular
complex, but is most probably a foreground object (Carpenter et al.\ 
\cite{car95}). Sh~259 is nearly circular around a central B1 star, and
lies at 8.3 kpc (Moffat et al.~\cite{mof79}). It is a thermal radio
continuum source (Wouterloot et al.~\cite{wou88}, Fich~\cite{fic93}) and
is bordered to the west by a half-ring of MSX emission. An MSX point
source lies in the middle of the arc. Near-IR objects are observed in
this direction. However, no H$_{2}$O maser has been detected (Wouterloot
et al.~\cite{wou93}).

\begin{figure*}
 \includegraphics[angle=270,width=85mm]{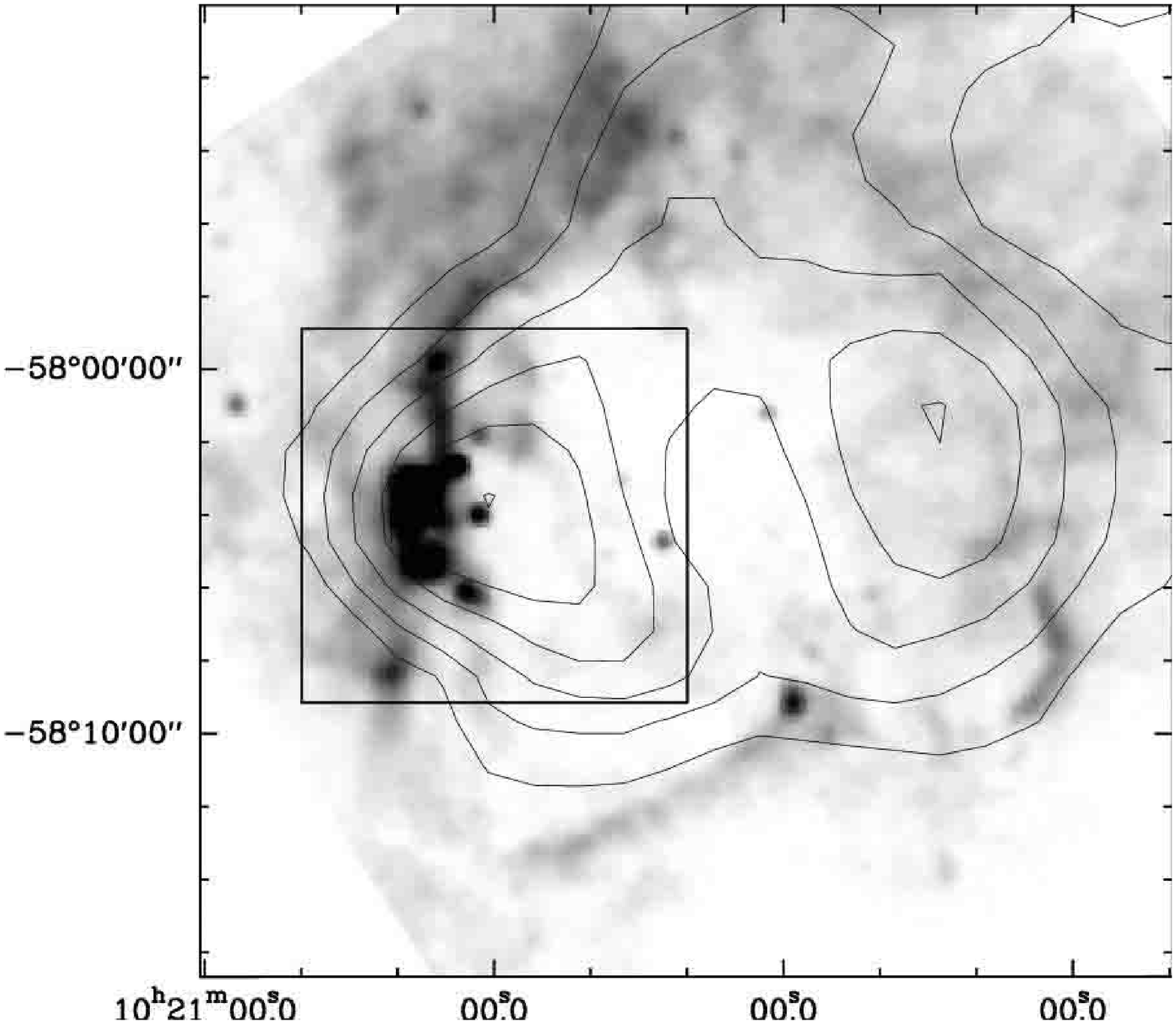}
 \includegraphics[angle=270,width=85mm]{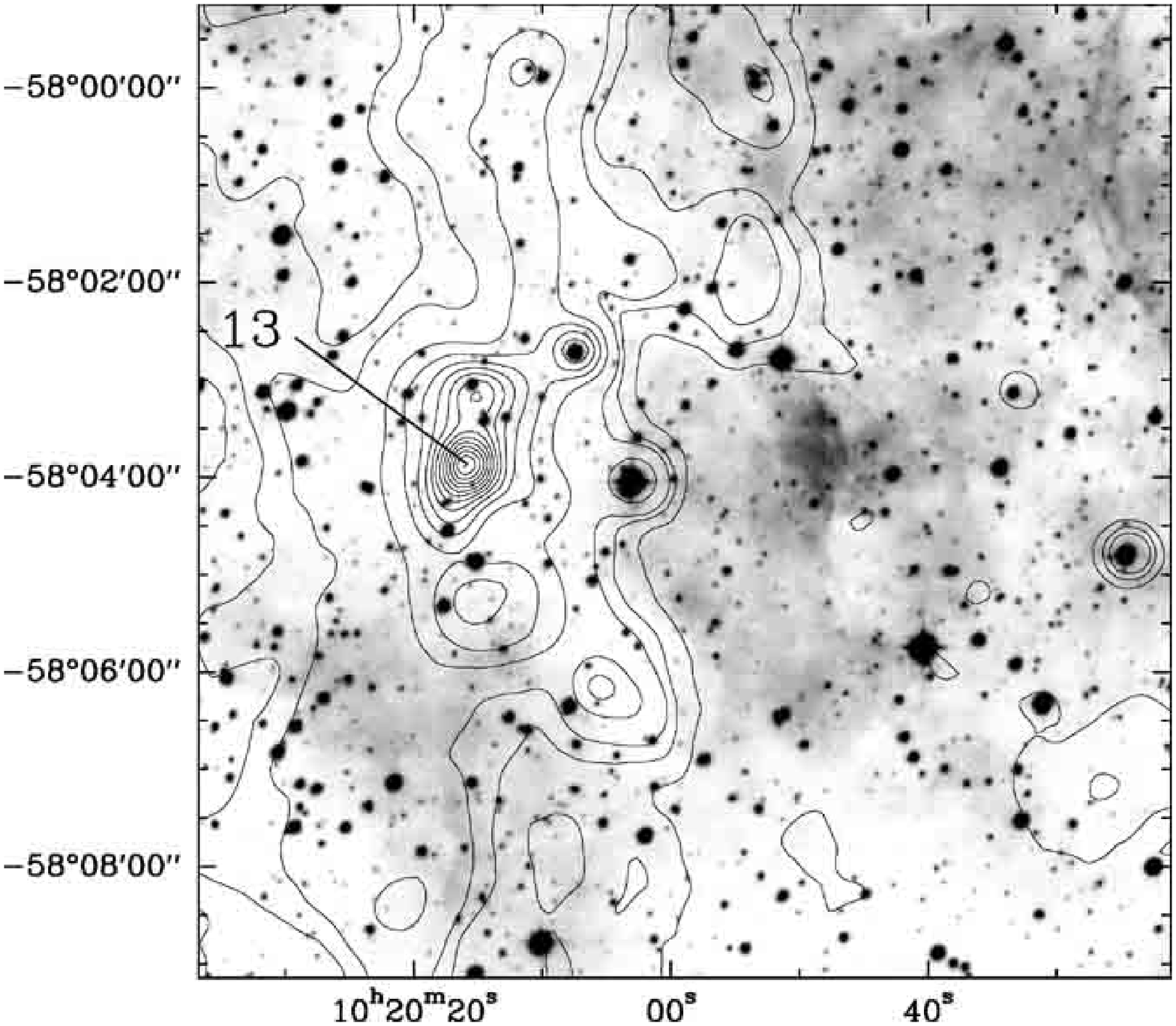}
 \caption{Dutra~45. {\it Left:} radio continuum emission at 4850~MHz (contours) 
superimposed on an MSX Band~A image; the square represents the field of 
the right image. {\it Right:} MSX Band A contours superimposed on an \Halpha\ 
image from SuperCOSMOS. The contour levels begin at $4\times10^{-6}$ and 
increase in  steps of $3\times10^{-6}$ W m$^{-2}$ sr$^{-1}$.}
\end{figure*}

\subsection{Southern hemisphere regions}

{\bf RCW~34} is a cometary shaped \HII\ region, located at 3.1~kpc and
excited by an O9Ib star (Russeil et al.\ \cite{rus03}, Avedisova \&
Kondratenko \cite{ave84}). MSX Band A emission surrounds the ionized
region with a bright MSX and IRAS point source located just in front of
the bright ionization front. Near-IR observations indicate that star
formation is observed at the border of the ionization front (Zavagno et
al.\ in preparation).

{\bf RCW~40} is an 8\arcmin\ diameter \HII\ region located at 1.84~kpc
and excited by an O9V star (Avedisova \&  Kondratenko \cite{ave84}). It
is surrounded by a complete ring of dust emission, with bright MSX and
IRAS point sources observed along the ring. The 1.2-mm continuum
observations reveal the presence of several bright condensations along
the ring (Zavagno et al.\ in preparation). This indicates that the
collect and collapse process may be at work here.

{\bf Dutra~45 - Gal 284.0-0.9} is located at 4.5~kpc 
(Russeil \cite{rus03}). An IR cluster (no.~45 in Dutra et al.\ \cite{dut03}) 
lies at the western edge of a faint optical nebulosity (Fig.~14 {\it Right}). 
A large-field image of this region (Fig.~14 {\it Left}) reveals the presence 
of a dust ring surrounding a radio emission region of 15$\arcmin$
diameter (Condon et al.~\cite{con93}, observations at 4850~MHz). 
The near-IR cluster that corresponds to the bright MSX point
source listed in Table~1 has been observed in the near IR. 
The images reveal the presence of a filamentary
structure that follows the optical ionization front (Zavagno et
al.\ in preparation). This
filament contains numerous bright and very red objects, indicating that
star formation is taking place there.

{\bf Dutra~46 - Gal 284.723+0.313} lies at a distance of 
6.2~kpc (Bronfman et al.~ \cite{bro96}). Faint
optical emission is observed, surrounded by
extended MSX Band A emission. Our near-IR observations of the MSX point
source reveal the presence of two very bright, red clusters. The
brighter one corresponds to no.~46 in Dutra et al.\
(\cite{dut03}). Star formation is ongoing at the border of the
ionized structure.

\begin{figure}
  \includegraphics[angle=270,width=85mm]{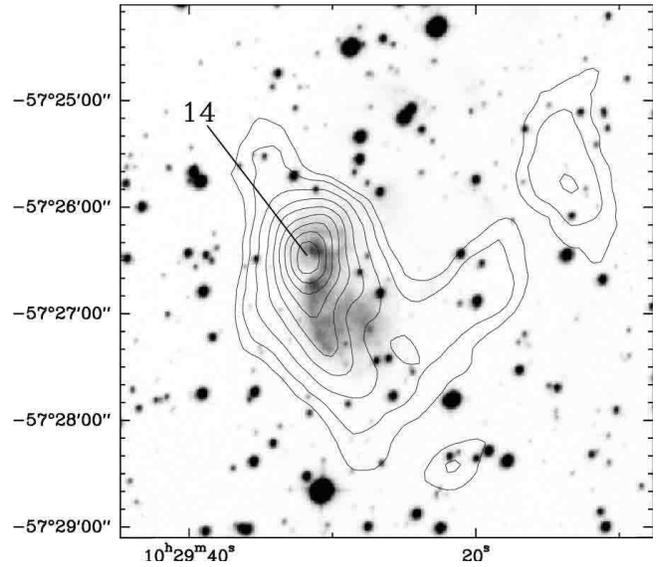}
\caption{Dutra~46. MSX Band A contours 
superimposed on an \Halpha\ image from SuperCOSMOS. The first contour levels 
are at $5\times10^{-6}$ and $7\times10^{-6}$ and then 
increase in  steps of $3\times10^{-6}$ W m$^{-2}$ sr$^{-1}$.}
\end{figure}

{\bf RCW~71} is located at 2.1~kpc and is excited by an O9.5V star 
(Russeil~\cite{rus03}, Avedisova \& Kondratenko \cite{ave84}). The
ionized region is surrounded by a complete ring of MSX Band A emission.
A bright MSX point source is observed on the ring, south-east of the
\HII\ region. The 1.2-mm continuum observations reveal the presence of five
condensations along the dust ring (Zavagno et al.\ in preparation),
indicating that star formation is taking place there, possibly 
by the collect and collapse mechanism.

\begin{figure}
  \includegraphics[angle=270,width=85mm]{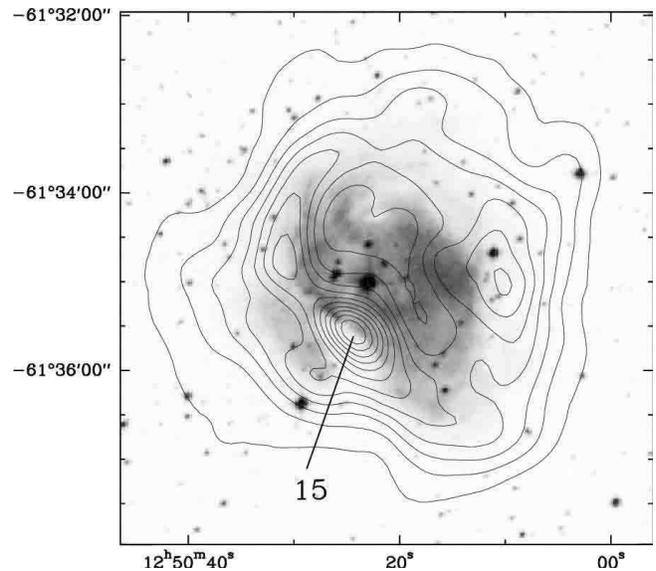}
  \caption{RCW~71. MSX Band A contours superimposed on
  a DSS-2 red image. The contour levels begin at $4\times10^{-6}$ and 
  increase in  steps of $2\times10^{-6}$ W m$^{-2}$ sr$^{-1}$.}
\end{figure}

{\bf RCW~79} is a bright optical \HII\ region of diameter
$\sim10\arcmin$, located at 4.2~kpc (Russeil~\cite{rus03}). It is
excited by a central 08V star and has been studied in detail by Cohen et al.\ 
(\cite{coh02}). This region is surrounded by MSX Band A emission
(Fig.~1). The MSX point source listed in Table~1 is double-peaked. A
compact \HII\ region (Cohen et al.\ \cite{coh02}) coincides with the
brightest peak. Our near-IR observations reveal the presence of a bright
cluster containing OB stars (Zavagno et al.\ in preparation). The other
peak is fainter, redder, and is associated with maser emission (Caswell
\cite{cas04}) suggesting that ongoing massive-star formation is taking
place. Moreover, our 1.2-mm continuum emission observations reveal the
presence of massive fragments distributed along the dust ring (Zavagno
et al.\ in preparation). These data strongly suggest that the collect
and collapse process is at work in this region.

\begin{figure}
  \includegraphics[angle=270,width=85mm]{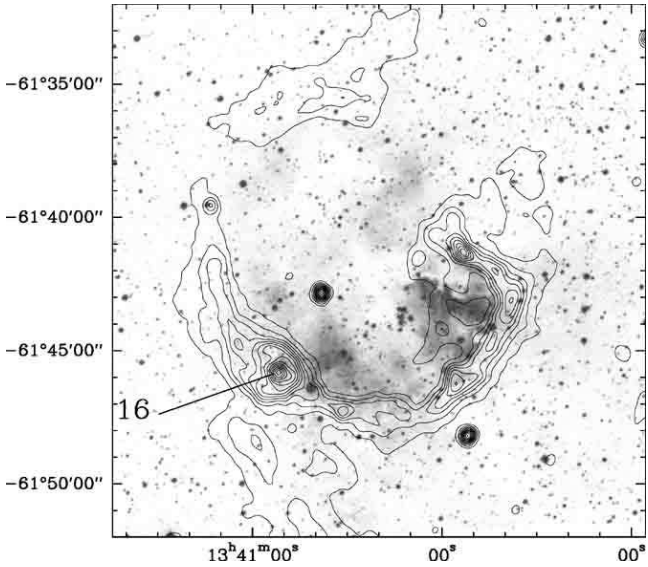}
  \caption{RCW~79. MSX Band A contours superimposed on
  an \Halpha\ image from SuperCOSMOS. The contour levels begin at $5\times10^{-6}$ and 
  increase in  steps of $3\times10^{-6}$ W m$^{-2}$ sr$^{-1}$.}
\end{figure}

{\bf RCW~82} is located at 2.9~kpc (Russeil \cite{rus03}), and has an
optical diameter of $\sim5\arcmin$. It is surrounded by a
nearly complete MSX Band A emission ring (Fig.~1). Two bright MSX point 
sources are observed along this ring, diametrically opposite
each other. Point source 17 (in Table~1) corresponds
to near-IR sources which are bright and red (Zavagno et al.\ in 
preparation). Bright, red stars are also observed in the direction 
of the point source 18 (2MASS $K_S$ survey). This
indicates that star formation is taking place in the dust
ring that surrounds RCW~82.

\begin{figure}
  \includegraphics[angle=270,width=85mm]{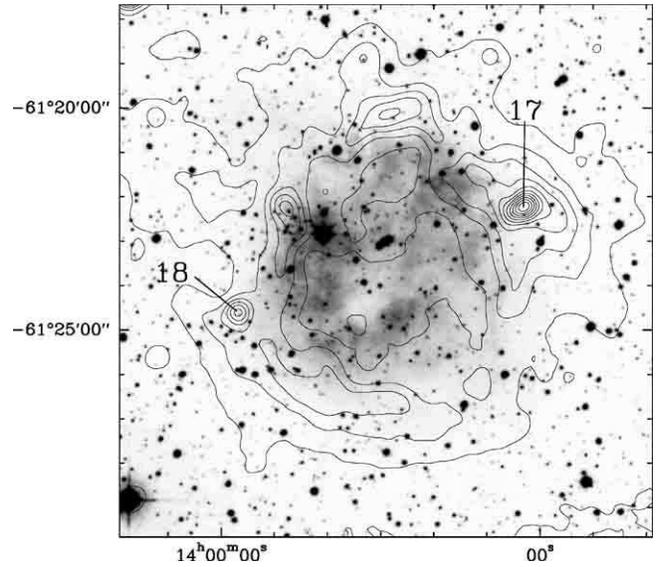}
  \caption{RCW~82. MSX Band A contours superimposed on
  an \Halpha\ image from SuperCOSMOS. The contour levels begin at $3\times10^{-6}$ and 
  increase in  steps of $2\times10^{-6}$ W m$^{-2}$ sr$^{-1}$.}
\end{figure}

{\bf RCW~120} is located at 1.2~kpc and is excited by an O6V star 
(Russeil~\cite{rus03}, Avedisova \& Kondratenko \cite{ave84}). Its
optical diameter is about $8\arcmin$. This region is almost completely
surrounded by a ring of MSX Band A emission (Fig.~1). Dust continuum
emission at 1.2~mm reveals the presence of five fragments distributed
along the ring (Zavagno et al.\ in preparation). One of them
corresponds to the MSX point source listed in Table~1. The dust ring is
well defined south of RCW~120, is elongated north-south, and is possibly
open at its northern extremity. The ionized gas seems to flow away from
the \HII\ region through this hole. A similar picture is given by RCW~79.
These two \HII\ regions have possibly evolved in a medium having a
density gradient, and are presently experiencing a champagne flow, after
the fragmentation of their surrounding dust rings. The presence of cold
dust condensations on the ring, along with the presence of near-IR objects
detected with 2MASS, indicates that the collect and collapse process may
be at work here.

\begin{figure}
  \includegraphics[angle=270,width=85mm]{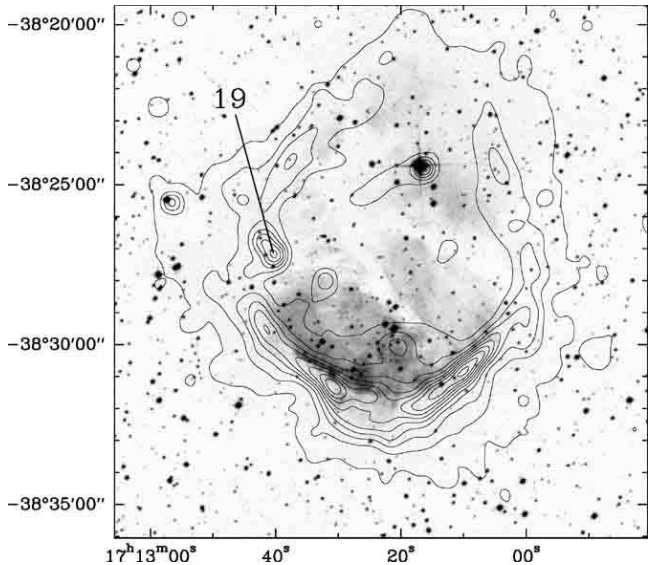}
  \caption{RCW~120. MSX Band A contours superimposed on
  a DSS-2 red image. The contour levels begin at $4\times10^{-6}$ and 
  increase in  steps of $3\times10^{-6}$ W m$^{-2}$ sr$^{-1}$.}
\end{figure}

\section{Discussion and conclusion\label{discconc}}

We have selected seventeen \HII\ regions with a very simple morphology.
These regions are relatively isolated, most of them being 
situated somewhat north or south of the Galactic plane. 
The survival of an annular structure around an ionized region may be
easier in `quiet' zones (our selection is deliberately biased toward
regions presenting a simple morphology, and this is favoured by
quiet surroundings).
 
We have observed most of these MSX point sources in the near IR, at the
CFH or at the ESO NTT, in order to determine their stellar contents.
The associated embedded clusters, when present, are possibly
second-generation clusters. Some of them -- those associated with Sh~104,
Sh~217 and RCW~79 -- contain massive stars exciting UC \HII\ regions. We
have shown that a few others -- those associated with Sh~212, Sh~241, 
Sh~259, RCW~34, Dutra~45 and Dutra~46 -- also probably contain massive
stars; this last point remains to be confirmed. The structure of these
clusters, if they are confirmed to be second-generation clusters --
hence young {\em relatively unevolved} clusters -- should give us information
about the way massive stars form. Are the massive stars
found in the very centres of the clusters, and do the clusters show
mass segregation? In some regions, such as Sh~212, only a bright {\em
isolated} source is observed. It seems very important, in the context
of massive-star formation, to find similar isolated objects, and to
estimate their masses and their evolutionary stages.

We have observed the molecular and dust environment of a few \HII\
regions at millimetre wavelengths. The molecular line emission in 
Sh~104, Sh~217 and Sh~219 was mapped at IRAM. The
continuum emission of cold dust, at 1.2~mm, was 
mapped with the SEST at ESO in RCW~40, RCW~71, RCW~79 and RCW~120,
and with MAMBO at IRAM in Sh~217, source Z, Sh~241 and Sh~259. These
observations are important because only the presence of a dense
molecular shell surrounding the ionized gas, or the presence of massive
fragments regularly spaced along the ionization front, can prove that
we are dealing with the collect and collapse process. This is
definitely the case for Sh~104 (Deharveng et al.~\cite{deh03b}), and
most probably the case for RCW~79 (Zavagno et al.\ in preparation). The
situation is unclear for Sh~217 and especially Sh~219, which are
partially surrounded by a low density ring of atomic (and not high
density molecular) hydrogen.

With our criteria we have selected two types of \HII\ regions. Some 
(Sh~104, RCW~40, RCW~79, RCW~82 and RCW~120) are surrounded by a nearly
complete ring of dust emission. Some others are low brightness \HII\
regions, surrounded by an incomplete ring of dust emission (source Z,
Sh~219, Sh ~241 and Sh~259). Star formation is presently taking place
at the periphery of Sh~241, as well as at the periphery of RCW~79, as
attested by the detection of H$_2$O masers at the
peripheries of both of these regions. It will be interesting to
determine if the collect and collapse process is at work in both types
of \HII\ regions, and if the same kind of stars are formed.

In a few cases one may wonder if the cluster observed at the 
periphery of the \HII\ region is not the primary source of 
ionization of the gas, the extended \HII\ region being a blister 
on the surface of the molecular cloud, with the ionized gas flowing away 
from the cloud. In the case of Sh~219 for example, there 
is some evidence for a `Champagne flow': the velocity 
of the ionized gas is $\sim 6$~km~s$^{-1}$ more negative than 
that of the molecular material (Deharveng et al.~\cite{deh03a}). However, the 
exciting star of Sh~219 is clearly identified, and lies at the centre 
of the \HII\ region and not in the cluster observed at its periphery. 
The regions Sh~104, Sh~212, Sh~217, Sh~241, Sh~259, RCW~40, RCW~79  
and RCW~120 are also excited by identified stars lying near 
their centres. Lopsided RCW~34 is possibly a 
blister \HII\ region, but its exciting star is identified
(Heydari-Malayeri~\cite{hey88}); it lies near the ionization front but 
is separated from the MSX point source. Nothing is known about the 
exciting stars of sources X, Y,  Z and Dutra~46.

The molecular or cold dust condensations observed at the periphery of a
given region contain objects in various evolutionary stages. For
example, around RCW~79 are found: i) a near-IR cluster associated with a
compact \HII\ region -- thus already a bit evolved; ii) a red MSX point
source, in the direction of which an H$_{2}$O maser is detected -- hence a
present site of star formation; iii) condensations with no associated
near- or mid-IR sources, thus, possibly, recently formed by fragmentation,
or containing class 0 objects. This indicates either that the fragmentation of
the compressed layer spreads over time, or that the fragments evolve at 
different rates -- a complication compared to the models.

In several regions, such as Sh~104, RCW~79 and RCW~120, the molecular or
cold dust condensations are almost all observed in the direction of the
dust ring, as if the fragmentation (and subsequent star formation) were
taking place in a preferential plane perpendicular to the line of sight. 
This is not
predicted by the collect and collapse model, where fragmentation occurs
in the spherical shell, at the periphery of the ionized region (so a few
condensations should be observed inside the ring). A new model which can
explain these observations is presently being developed by A.~Whitworth
(private communication).

\begin{acknowledgements}

We would like to thank R.~Cautain for his development of procedures to
transform the images, and Y.~Brand, B.~Lefloch and F.~Massi for their
collaboration in this long term project. This research has made use of
the Simbad astronomical database operated at CDS, Strasbourg, France,
and of the interactive sky atlas Aladin (Bonnarel et al.~\cite{bon00}).
This publication uses data products from the Midcourse Space
EXperiment, from the Two Micron All Sky Survey and from the InfraRed
Astronomical Satellite; for these we have used the NASA/IPAC Infrared
Science Archive, which is operated by the Jet Propulsion Laboratory,
California Institute of Technology, under contract with the National
Aeronautics and Space Administration. We have also used the
SuperCOSMOS, NVSS and GB6 surveys.

\end{acknowledgements}


{}


\end{document}